\pdfoutput=1
\documentclass[footinbib,preprint,prd,tightenlines, superscriptaddress]{revtex4-1}

\usepackage{graphicx} 
\usepackage{dcolumn}  
\usepackage{colordvi}
\usepackage{color}
\usepackage{epstopdf}
\usepackage{csquotes}
\usepackage{subfig}
\usepackage{caption}
\usepackage{amssymb}
\usepackage{amsmath}
\usepackage{url}
\usepackage{cleveref}
\usepackage{tabularx}
\usepackage{booktabs}
\usepackage[italic]{hepnames}
\graphicspath{{figures/}}
\usepackage{siunitx}
\DeclareSIUnit \clight {\text{\ensuremath {c}}}
\usepackage{grffile}
\usepackage{orcidlink}

\renewcommand{\arraystretch}{1.1}

\renewcommand{\Pfz}{\HepParticle{f}{0}{}{(980)}}

\input belle2sym.tex

\def\Btag        {\ensuremath{B_{\mathrm{tag}}}\xspace}

\def\Bsig        {\ensuremath{B_{\mathrm{sig}}}\xspace}

\def\X           {\ensuremath{X}\xspace}

\def\Xu          {\ensuremath{X_{u}}\xspace}

\def \Y       {\ensuremath{\Upsilon(4S)}\xspace}

\newcommand{\mpipi}{\ensuremath{M(\pi\pi})\xspace}
\newcommand{\mmisssq}{\ensuremath{M_{\mathrm{miss}}^2}}

\newcommand\bfrhoc{\ensuremath{\mathcal{B}}(\ensuremath{B^0\to\rho^-\ell^+\nu_\ell})}
\newcommand\bfrhoz{\ensuremath{\mathcal{B}}(\ensuremath{B^+\to\rho^0\ell^+\nu_\ell})}

\newcommand{\resultrhoc}{\ensuremath{(4.12 \pm 0.64 (\mathrm{stat}) \pm 1.16 (\mathrm{syst})) \times 10^{-4}}}
\newcommand{\resultrhoz}{\ensuremath{(1.77 \pm 0.23 (\mathrm{stat}) \pm 0.36 (\mathrm{syst})) \times 10^{-4}}}
\newcommand{\resultrhocstat}{\ensuremath{(4.12 \pm 0.64 (\mathrm{stat})) \times 10^{-4}}}
\newcommand{\resultrhozstat}{\ensuremath{(1.77 \pm 0.23 (\mathrm{stat})) \times 10^{-4}}}

\newcommand{\PXu}{\HepParticle{X}{u}{}}
\newcommand{\PXc}{\HepParticle{X}{c}{}}

\newcommand{\Pfn}{\HepParticle{f}{n}{}}

\newcommand{\Btorholnu}{\HepProcess{\PB\to\Prho\Plepton\Pnulepton}}
\newcommand{\Btorhoplus}{\HepProcess{\PBzero\to\Prhominus\Pleptonplus\Pnulepton}}
\newcommand{\Btorhozero}{\HepProcess{\PBplus\to\Prhozero\Pleptonplus\Pnulepton}}
\newcommand{\BtoXulnu}{\HepProcess{\PB\to\PXu\Plepton\Pnulepton}}
\newcommand{\Btopizero}{\HepProcess{\PBplus\to\Ppizero\Pleptonplus\Pnulepton}}
\newcommand{\Btopiplus}{\HepProcess{\PBzero\to\Ppiminus\Pleptonplus\Pnulepton}}
\newcommand{\Btopipilnu}{\HepProcess{\PB\to(\Ppi\Ppi)\Pleptonplus\Pnulepton}}

\def\lint {189.9 \invfb}

\begin{document}

\vspace*{-3\baselineskip}
\resizebox{!}{3cm}{\includegraphics{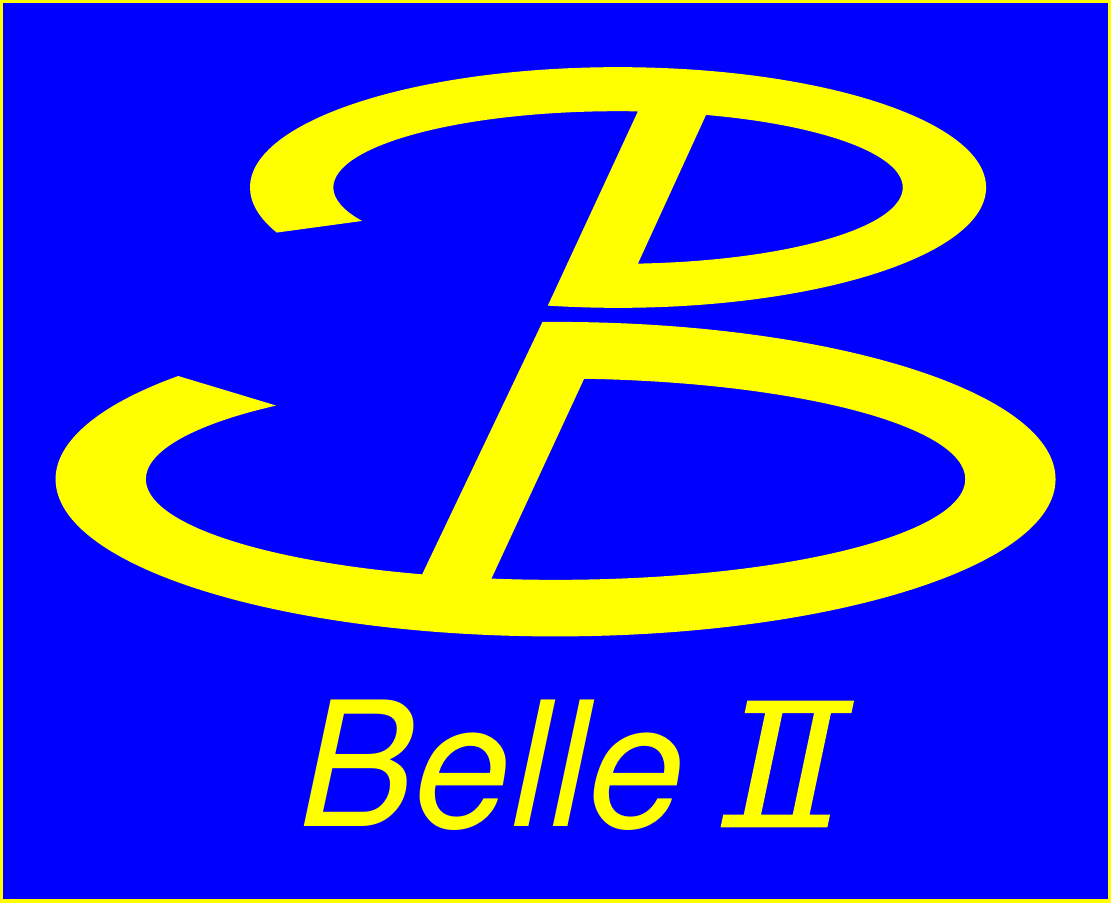}}

\vspace*{-5\baselineskip}
\begin{flushright}
\vspace{1em}
\today\\
{BELLE2-CONF-PH-2022-029}

\end{flushright}

\vspace{2em}
\title {\LARGE Reconstruction of \Btorholnu{} decays identified using hadronic decays of the recoil \PB{} meson in 2019 -- 2021 Belle II data\\
\vspace{1em}
}

  \author{F.~Abudin{\'e}n\,\orcidlink{0000-0002-6737-3528}} 
  \author{I.~Adachi\,\orcidlink{0000-0003-2287-0173}} 
  \author{K.~Adamczyk\,\orcidlink{0000-0001-6208-0876}} 
  \author{L.~Aggarwal\,\orcidlink{0000-0002-0909-7537}} 
  \author{P.~Ahlburg\,\orcidlink{0000-0002-9832-7604}} 
  \author{H.~Ahmed\,\orcidlink{0000-0003-3976-7498}} 
  \author{J.~K.~Ahn\,\orcidlink{0000-0002-5795-2243}} 
  \author{H.~Aihara\,\orcidlink{0000-0002-1907-5964}} 
  \author{N.~Akopov\,\orcidlink{0000-0002-4425-2096}} 
  \author{A.~Aloisio\,\orcidlink{0000-0002-3883-6693}} 
% ...
\collaboration{The Belle II Collaboration}

\begin{abstract}
We present results on the semileptonic decays \Btorhoplus{} and \Btorhozero{} in a sample corresponding to \lint of Belle II data at the SuperKEKB $\Pelectron\APelectron$ collider.
Signal decays are identified using full reconstruction of the recoil \PB{} meson in hadronic final states.
We determine the total branching fractions via fits to the distributions of the square of the \enquote{missing} mass in the event and the dipion mass in the signal candidate and find ${\bfrhoc = \resultrhoc}$ and ${\bfrhoz = \resultrhoz}$ where the dominant systematic uncertainty comes from modeling the nonresonant \Btopipilnu{} contribution. 
\end{abstract}
\maketitle

\section{Introduction}

The Belle II detector has been collecting physics data from electron-positron collisions at the SuperKEKB collider since 2019.
In this document, we present results on the decays \Btorhoplus{} and \Btorhozero{}, where $\ell = e$, $\mu$, \footnote{Charge conjugate processes are implied for all quoted decays of \PB{} mesons throughout this document.} via hadronic $B$-tagging, in a sample corresponding to \lint of Belle II data .
These decays offer access to determinations of the magnitude of the Cabibbo-Kobayashi-Maskawa (CKM) matrix element $|V_{\mathrm{ub}}|$, independent from the modes \Btopiplus{} and \Btopizero{}, which are commonly used for this purpose. 
Here $B$-tagging refers to the reconstruction, in hadronic final states, of the second \PB{} meson in a \HepProcess{\PUpsilonFourS\to\PB\APB} decay and uses the Full~Event~Interpretation (FEI) algorithm \cite{Keck:2018lcd}. 
Both of the two most precise previous measurements of the branching fraction \bfrhoc{} and the two most precise previous measurements of the branching fraction \bfrhoz{} are in significant tension with each other~\cite{BaBar:2010efp, Sibidanov:2013sb}.
An inclusive measurement of \HepProcess{\PB\to\Ppi\Ppi\Plepton\Pnulepton}, reconstructing the same final state as \Btorhozero{}, has been reported in Ref. \cite{beleno2021} and allows for the presence of a nonresonant component in the dipion mass, which might explain this discrepancy.

The results presented here supersede the preliminary Belle~II results based on a \SI[per-mode=reciprocal]{62.8}{\per\femto\barn} data set~\cite{prelim_xulnu_2021}.
Compared to the previous result, which employed a single variable for signal extraction, the larger data set allows the use of a two-dimensional signal region that provides additional discrimination between the signal and nonresonant \HepProcess{\PB\to\Ppi\Ppi\Plepton\Pnulepton} background.


\section{The Belle II Detector}
The Belle~II detector is described in detail in Ref. \cite{Abe:2010sj}. 
The $z$-axis of the Belle~II detector is defined as the symmetry axis of the solenoid, and the positive direction is approximately given by the electron-beam direction. The polar angle $\theta$, as well as the longitudinal and the transverse directions, are defined with respect to the $z$-axis.
The innermost layers are known collectively as the vertex detector and are dedicated to the precise determination of particle decay vertices. The vertex detector is composed of two layers of silicon pixel sensors in which the outer layer is incomplete and covers 15\% of the azimuthal acceptance. These pixel sensor layers are surrounded by four layers of silicon strip detectors.
A central drift chamber (CDC) surrounds the vertex detector, encompassing the barrel region ($17^\circ < \theta < 150^\circ$) of the detector, and is primarily responsible for the reconstruction of charged particle trajectories (tracks).
Particle identification is provided by two independent Cherenkov-imaging systems, the time-of-propagation detector system and the aerogel ring-imaging Cherenkov detector, located in the barrel and forward endcap ($14^\circ < \theta < 30^\circ$) regions of the detector surrounding the CDC. The electromagnetic calorimeter (ECL) consists of forward, backward and barrel regions. The ECL encloses all of the sub-detector systems described above and is used primarily for the determination of the energies of electrons and photons.
A superconducting solenoid surrounds the inner components and provides a \SI{1.5}{\tesla} axial magnetic field. Finally, outermost subdetectors detect $K^0_L$ mesons and muons.

\section{Data sets}
\label{sec:data sets}

The data studied for this analysis correspond to an integrated luminosity of \lint collected at collision energies close to the mass of the \PUpsilonFourS{} resonance. In addition, a smaller data set was collected at beam energies \SI{60}{\MeV} below the \PUpsilonFourS{} mass, corresponding to \SI[per-mode=reciprocal]{9.5}{\per\femto\barn}.
To derive the reconstruction efficiencies of signal events, study the background composition, and define templates used in the signal extraction, simulated Monte Carlo (MC) samples corresponding to a total integrated luminosity of 1\invab are used. These contain events with decays of pairs of charged or neutral $B$ mesons, as well as continuum \epem \to \qqbar processes where $q$ indicates a $u$, $d$, $s$, or $c$ quark.
Beam-background effects including beam scattering and radiative processes are simulated separately and overlaid in each event to represent the conditions in the detector.

In addition to these generic MC samples, dedicated samples in which one \PB{} meson decays as \BtoXulnu{}, where $X_u$ is a hadronic system resulting from the quark-flavor transition $b \to u$, are used to model signal decays and related backgrounds. The $X_u$ system includes both resonant and nonresonant contributions using the hybrid modeling technique of Ref. \cite{Ramirez:1990db}, which is briefly described here.
In addition to the resonant decays \HepProcess{\PB\to\Ppi\Pleptonplus\Pnulepton}, \HepProcess{\PB\to\Prho\Pleptonplus\Pnulepton}, \HepProcess{\PBplus\to\Peta\Pleptonplus\Pnulepton}, \HepProcess{\PBplus\to\Petaprime\Pleptonplus\Pnulepton}, and \HepProcess{\PBplus\to\Pomega\Pleptonplus\Pnulepton} with well-measured branching fractions, contributions from two unobserved resonant decays are included in the simulated samples.
The branching fractions for these decays are assumed to be ${\mathcal{B}(\HepProcess{\PBplus\to\Pfz\Pleptonplus\Pnulepton}) = 1 \times 10^{-5}}$ and ${\mathcal{B}(\HepProcess{\PBplus\to\Pfii\Pleptonplus\Pnulepton}) = 1.62 \times 10^{-4}}$, where the latter is taken from~\cite{Sibidanov:2013sb}. For the former decay, no such estimates exist. 
The branching fraction of the decay \HepProcess{\Pfii\to\Ppiplus\Ppiminus} is assumed to be 0.57 according to Ref.~\cite{Zyla:2020zbs}; for $\mathcal{B}$(\HepProcess{\Pfz\to\Ppiplus\Ppiminus}) a branching fraction of 0.5 is assumed, which is consistent with Ref.~\cite{PhysRevD.72.092002}.

In the generic MC samples used in Belle~II, the two \HepProcess{\PBplus\to\Pfn\Pleptonplus\Pnulepton} decay modes mentioned above are not included. When adding them, any model involving the same final states in a nonresonant amplitude must be rescaled to maintain the total rate. 
In addition to the resonant (R) exclusive decays described above, a second sample containing 50 million nonresonant (NR) events corresponding to an inclusive component, described using the BLNP heavy-quark-effective-theory-based model \cite{Lange:2005ll}, is generated. The hadronic system \PXu{} from this inclusive model is then fragmented using the \texttt{PYTHIA} software package~\cite{Sjostrand:2014zea}. The inclusive and exclusive samples are combined and the eFFORT tool \cite{markus_prim_2020_3965699} is used to calculate an event-wise weight $w_i$ as a step-wise function of the generated lepton energy in the $B$-frame, $E^B_\ell$; the squared four-momentum transfer to the leptonic system, $q^2$; and the mass of the hadronic system containing an up-quark, $M_X$; such that the relation $H_i = R_i + w_i {NR}_i$ holds. 
The \BtoXulnu{} events from the generic 1\invab MC samples are replaced with the equivalent amount of this \enquote{hybrid MC}.

\section{Full Event Interpretation}
\label{sec:FEISkim}
The Full Event Interpretation (FEI) algorithm \cite{Keck:2018lcd} uses machine learning to identify whether a collision produced a $B\overline{B}$ pair (\enquote{tag the event}) and infer the kinematic properties of the signal by reconstructing the second \PB{} meson (\enquote{\Btag{}}) in the event.
FEI can identify \Btag{} decays in both hadronic and semileptonic final states, reconstructing $B$ mesons in more than 4000 individual decay chains. The algorithm uses the FastBDT software package to train a series of multivariate classifiers for each tagging channel via a number of stochastic gradient-boosted decision trees \cite{Keck:2016tk}. The training is performed in a hierarchical manner with final-state particles being reconstructed first from detector information. The decay channels are then built up from these particles and used to reconstruct \PB{} mesons. For each $B$-meson tag candidate reconstructed by the FEI algorithm, a value of the final multivariate classifier output, the \texttt{SignalProbability}, is assigned. The \texttt{SignalProbability} is distributed between zero and one, representing candidates identified as being background-like and signal-like, respectively.

For each event reconstructed using the hadronic FEI, three or more tracks are required, as the vast majority of $B$-meson decay chains reconstructable by hadronic FEI include at least three charged particles. Requirements are placed on the track impact parameters as defined in Ref. \cite{belle2tracking:2021} to ensure close proximity to the interaction point (IP), with the distances from the nominal center of the detector along the $z$-axis and in the transverse plane satisfying $|z_0| < 2.0\,\rm cm$ and $|d_0| < 0.5\,\rm cm$, respectively.
A minimum threshold $p_t >$ 0.1 GeV/$c$ is placed on the charged particle's transverse momentum. Similar requirements are applied to the localized energy deposits (clusters) in the ECL, with at least three signals satisfying a minimum deposited energy threshold $E > 0.1$ GeV within the polar angle acceptance of the CDC, 0.3 $< \theta <$ 2.6 rad.
The total detected energy in each event is required to be at least 4 GeV. The total energy deposited in the ECL is restricted to the range $2 < E_{\rm ECL} < 7\,{\rm GeV}$, however, to suppress events with an excess of energy due to beam background.

Application of the FEI algorithm typically results in up to 20 $B_{\mathrm{tag}}$ candidates per event. The number of these candidates is reduced with selections on the beam-constrained mass, $M_{\mathrm{bc}}$, and energy difference, $\Delta E$,

\begin{equation*}
M_{\mathrm{bc}} = \sqrt{E_{\mathrm{beam}}^{2} - \vec{p}^{\,2}_{B_{\mathrm{tag}}}}\hspace{0.5em},  \hspace{3em} \Delta E = E_{B_{\mathrm{tag}}} - E_{\mathrm{beam}}
\end{equation*}
where $E_{\mathrm{beam}}$ is the beam energy in the center-of-mass system (cms) and $\vec{p}_{B_{\mathrm{tag}}}$ and $E_{B_{\mathrm{tag}}}$ are the  $B_{\mathrm{tag}}$ momentum and energy in the cms, respectively. The criteria applied are $M_{\mathrm{bc}} > 5.27$ GeV/$c^2$ and $|\Delta E| < 0.2$ GeV.

Finally, a loose requirement on the $B_{\mathrm{tag}}$ classifier output,  \texttt{SignalProbability} $> 0.001$, provides further background rejection with minimal signal loss. 
The $B_{\mathrm{tag}}$ candidate having the highest value of the \texttt{SignalProbability} classifier output is retained in each event.

As the multivariate classifiers used in the FEI are trained using simulated data, the efficiency of the tagging differs between recorded and simulated data. An independent analysis of inclusive semileptonic decays is performed using the momentum of the lepton to determine corrections as detailed in Ref.~\cite{Sutcliffe:2020cn}. To test the validity of the calibration factor and the associated systematic uncertainties for \Btorholnu{} decays, the normalization is evaluated in a sideband and, if necessary, extrapolated to the signal region as detailed in Section~\ref{sec:corr}. 

\section{Signal Selection}
\label{sec:selections}

The distribution of the square of the missing mass, $M_{\mathrm{miss}}^2$, and the invariant mass of the dipion system \mpipi{} are the variables chosen for the determination of the signal yields. We define the four-momentum of the signal \PB{} meson \Bsig{} in the cms as follows,

\begin{equation*}
    p_{B_{\mathrm{sig}}} \equiv  (E_{B_{\mathrm{sig}}}, \vec{p}_{B_{\mathrm{sig}}}) = \left(\frac{m_{\Upsilon(4S)}}{2}, -\vec{p}_{B_{\mathrm{tag}}}\right),
\end{equation*}
where $m_{\Upsilon(4S)}$ is the known $\Upsilon$(4S) mass \cite{Zyla:2020zbs}. We then define the missing four-momentum as

\begin{equation*}
    p_{\mathrm{miss}} \equiv (E_{\mathrm{miss}}, \vec{p}_{\mathrm{miss}}) =  p_{B_{\mathrm{sig}}} - p_Y,
\end{equation*}
where $Y$ represents the combined lepton--$\rho$-meson system. The square of the missing momentum is $M_{\mathrm{miss}}^2 \equiv p_{\mathrm{miss}}^2$.

The event selection closely follows that from a 2013 study of exclusive, hadronically-tagged $B \to \X_u \ell \nu_\ell$ decays reconstructed in the full 711\invfb Belle data set~\cite{Sibidanov:2013sb}. All selections are applied in addition to the hadronic FEI skim criteria described in the previous section.

\subsection{Selections on final state particles}

Tracks are required to have longitudinal axis and transverse-plane distances from the IP of $|dz| < 5\,{\rm cm}$ and $dr < 2$ cm, respectively, to suppress mismeasured tracks. 
Charged-particle momenta are multiplied by a factor to correct for momentum-scale differences between data and MC. 

Electrons and muons are reconstructed based on charged particles that meet specific particle-identification criteria.
Only leptons within the acceptance of the CDC are selected with a lab-frame momentum greater than $p_{\mathrm{lab}} > 0.4$ GeV/$c$. Electrons and muons are identified through selection criteria on the particle-identification variables. These variables describe the probability that each species of charged particle generates the observed particle-identification signal, and use information from all the individual subdetectors except the SVD.
Electron and muon candidates are each required to have an identification probability above $0.9$ as assigned by the appropriate reconstruction algorithm.
In \PB{}-meson decays and within the selected detector and momentum regions, the particle identification requirements select electrons with an efficiency of $(86.0 \pm 0.4)\%$ with pion mis-identification rates of $(0.41 \pm 0.02)\%$.
Muons are selected with an efficiency of $(88.5 \pm 0.4)\%$ with pion mis-identification rates of $(7.33 \pm 0.02)\%$. 

The four-momenta of the reconstructed electrons are corrected in order to account for bremsstrahlung radiation. 
Any energy deposit in the ECL not associated with a track and above a given energy is considered a bremsstrahlung photon if it is detected within a given angle from a reconstructed electron candidate. In this case, the four-momentum of the photon is added to that of the electron, and the photon is excluded from the rest of the event. 
To maximize the performance of this method, the electron candidates are separated into three momentum regions, from 0.4 to 0.6 GeV/$c$, from 0.6 to 1.0 GeV/$c$, and above 1.0 GeV/$c$. In each region, individual threshold values in angle and energy are determined by minimizing the root-mean-square deviation between the reconstructed and generated electron momenta in simulation.
Finally, a single lepton is retained in each event with the highest value of the lepton identification probability.

For the charged pions, similar impact parameter criteria are applied as those for the leptons, with $dr < 2$ cm and $|dz| < 4$ cm. Charged pions are required to lie within the CDC acceptance.
A pion-identification criterion is applied with an efficiency of about 89\% for correctly identified pions and a rejection rate of about 85\% for misidentified pions in \PB{} meson decays.

\subsection{Selections on recombined particles}

In reconstructing neutral pions, different thresholds on the final-state photon energies are required, depending on the polar angle of the candidate photon. These requirements are $E > 0.080$ GeV for the forward end-cap, $E > 0.030$ GeV for the barrel region, and $E > 0.060$ GeV for the backward end-cap. The diphoton mass is required to be within $0.120$ to $0.145$~$\text{GeV}/c^2$. Additionally, a selection cos$\psi(\gamma\gamma) > 0.4$ on the cosine of the lab-frame opening angle of the $\pi^0$ photons is applied in order to reject backgrounds.
A correction is applied to scale the energies of the photons assigned to the signal $\pi^0$ candidates to account for known photon-energy biases.

To reconstruct \Prho{} mesons, either two oppositely charged pions or a charged pion and a neutral pion are combined. The invariant mass of this pion pair \mpipi{} is required to be between 0.33 GeV/$c^2$ and 1.22 GeV/$c^2$ for charged \Prho{} mesons  and between 0.27 GeV/$c^2$ and 1.36 GeV/$c^2$ for neutral \Prho{} mesons to reject contributions from nonresonant $B \to \pi \pi \ell \nu_\ell$ decays while minimizing the impact on the signal yield. If multiple charged \Prho{} meson candidates are reconstructed, the candidate with the highest energy in the center-of-mass frame is chosen.
To test whether the two charged pions originate from the same production vertex, a vertex fit using the \texttt{TreeFitter} package \cite{KROHN2020164269} is carried out and \Prhozero candidates with failed vertex fits are rejected. If the vertex fit is successful, the momenta of the \Prho{} meson candidates are updated according to the fit result. However, the vertex fit does not modify \mpipi{}. 

The four-momenta of the reconstructed meson and lepton in the cms are combined into the pseudoparticle $Y$. The angle between the momenta of this $Y$ and the nominal signal \PB{} meson as calculated from the known beam energy and the \PB{} meson mass is then used to select correctly reconstructed events.
The cosine of this angle, $\mathrm{cos}(\theta_{BY})$, is defined as
\begin{equation*}
\mathrm{cos}(\theta_{BY}) = \frac{2E_{\mathrm{beam}}E_Y - m_{B_{\mathrm{sig}}}^2 - m_Y^2}{2|\vec{p}_{B_{\mathrm{sig}}}||\vec{p}_Y|},
\end{equation*}
where $m_{B_{\mathrm{sig}}}$ is the mass of the signal \PB{} meson and $E_Y$, $m_Y$, and $\vec{p}_Y$ are the energy, mass, and momentum of the $Y$ system, respectively. A value of $\lvert \mathrm{cos}(\theta_{BY}) \rvert < 1$ is expected only if a single neutrino is missing in the reconstruction. However, this requirement is loosened to $ -3.0 <  \mathrm{cos}(\theta_{BY}) < 1.1$ to account for resolution effects and to avoid introducing potential bias in the background $M_{\mathrm{miss}}^2$ distributions. 

To test whether the reconstructed leptons and hadrons in the $Y$ system originate from the same vertex, a second vertex fit is carried out and $B$-meson candidates in which the fit fails are rejected. To limit the number of degrees of freedom in this fit, the mass of the \Ppizero{} candidate in the \HepProcess{\PB\to\Prhominus\Plepton\Pnulepton} reconstruction is constrained to the known value. Again, the particle momenta are updated according to the fit results. 

\subsection{Event-level selections and continuum suppression}

A minimum threshold on the missing energy in the event, $E_{\mathrm{miss}} > \SI{0.1}{\GeV}$, accounts for the neutrino. All charged particles and calorimeter energy depositions without associated charged particles remaining after the full reconstruction are combined into a single system known as the rest-of-event. Events in which additional tracks satisfying the conditions $dr < 2$ cm, $|dz| < 4$ cm and $p_t > 0.2$ GeV/$c$ remain after the reconstruction are excluded. For the remaining ECL clusters satisfying $E > 0.08$ GeV, $E > 0.03$ GeV, and $E > 0.06$ GeV for the forward end-cap, barrel and, backward end-cap regions, respectively, the energies are summed. This extra energy is required to be less than \SI{1.2}{\GeV} for $B\to\rho\ell\nu_\ell$ candidates.

In order to suppress events from the continuum background, a multivariate classifier is used. Several variables including the second normalized Fox-Wolfram moment, $B$-meson thrust angles and magnitudes, CleoCones~\cite{Cleo:1996}, modified Fox-Wolfram moments~\cite{Bfactories:2014}, and the beam-constrained mass of the \Btag{} are combined into a boosted-decision-tree classifier \cite{Keck:2016tk}. Out of the 64 available variables, the 61 with the highest classification power are selected.
The optimal classifier requirement is chosen by maximising the figure-of-merit
\begin{equation*}
	\lambda_{FOM} = \frac{N_{sig}}{\sqrt{N_{sig} + N_{bkg}}}
\end{equation*}
where $N_{sig}$ and $N_{bkg}$ are the numbers of simulated signal and background events in the signal region.
The resulting continuum background suppression  in the $B^+ \to \rho^0 \ell \nu_\ell$ reconstruction is 94.8\% while 94.4\% of signal events are retained. In the $B^0 \to \rho^+ \ell \nu_\ell$ reconstruction, 93.4\% of continuum events are rejected and 93.1\% of signal events are retained.

\begin{figure}[h!]
\begin{center}
\includegraphics[width=0.49\textwidth]{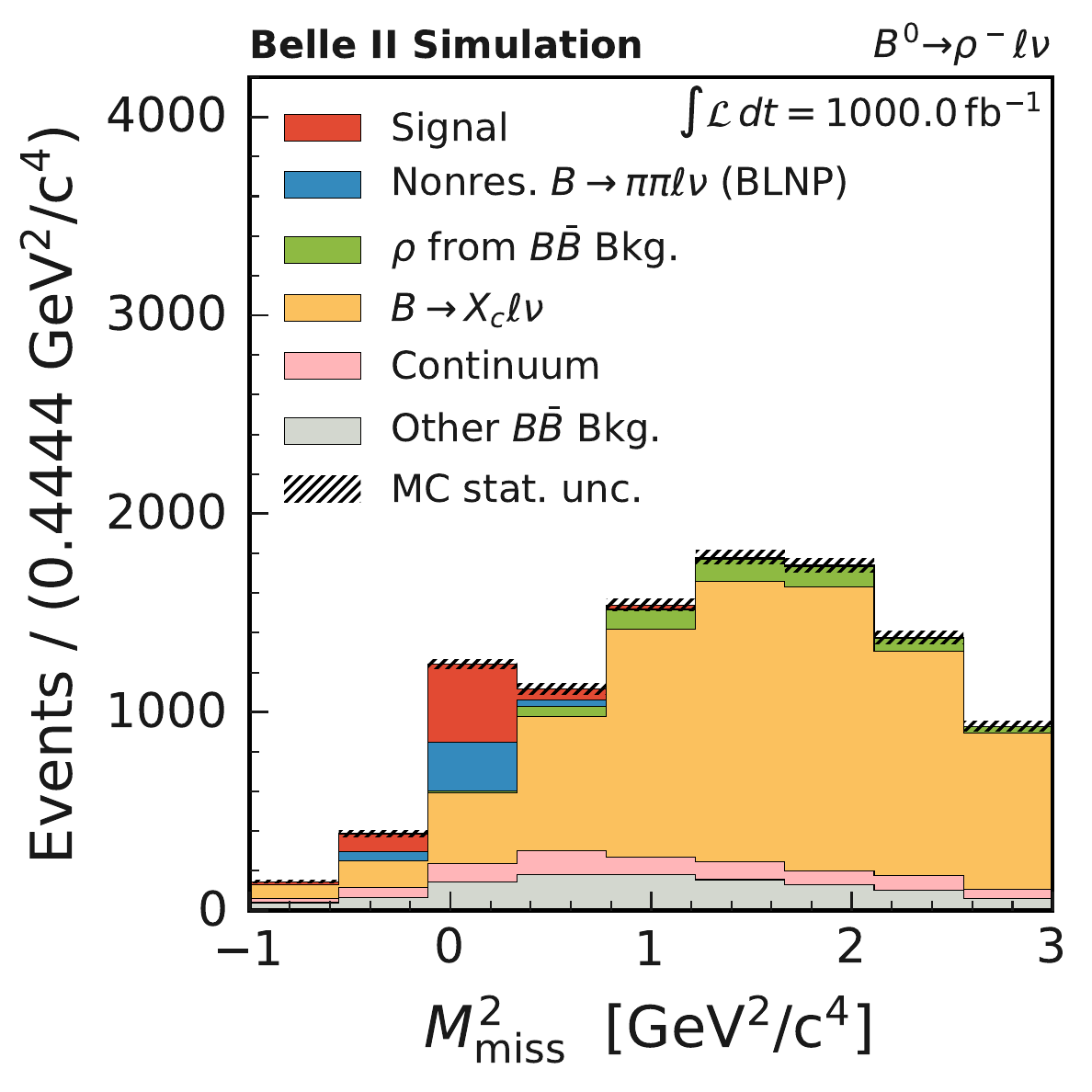}
\includegraphics[width=0.49\textwidth]{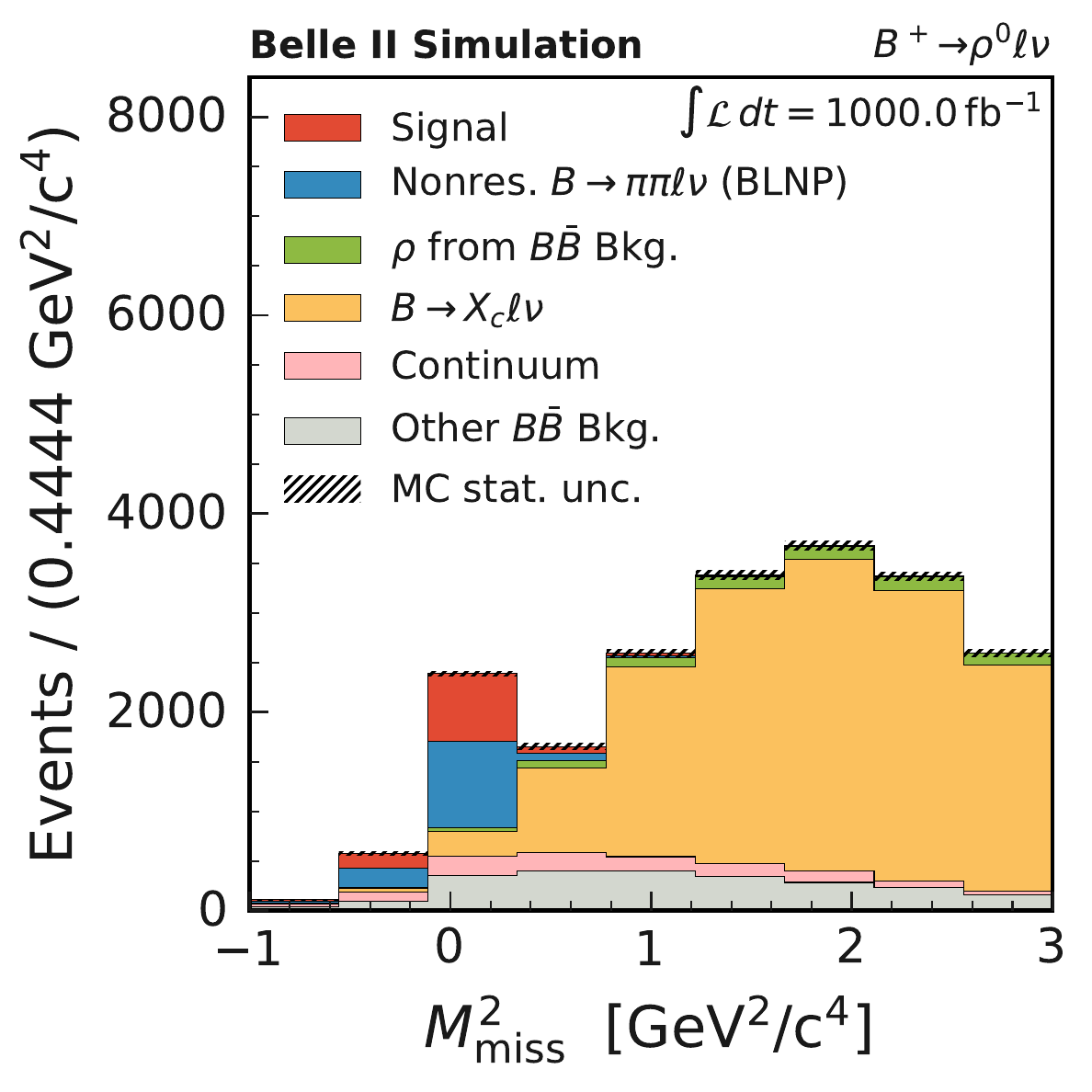}
\includegraphics[width=0.49\textwidth]{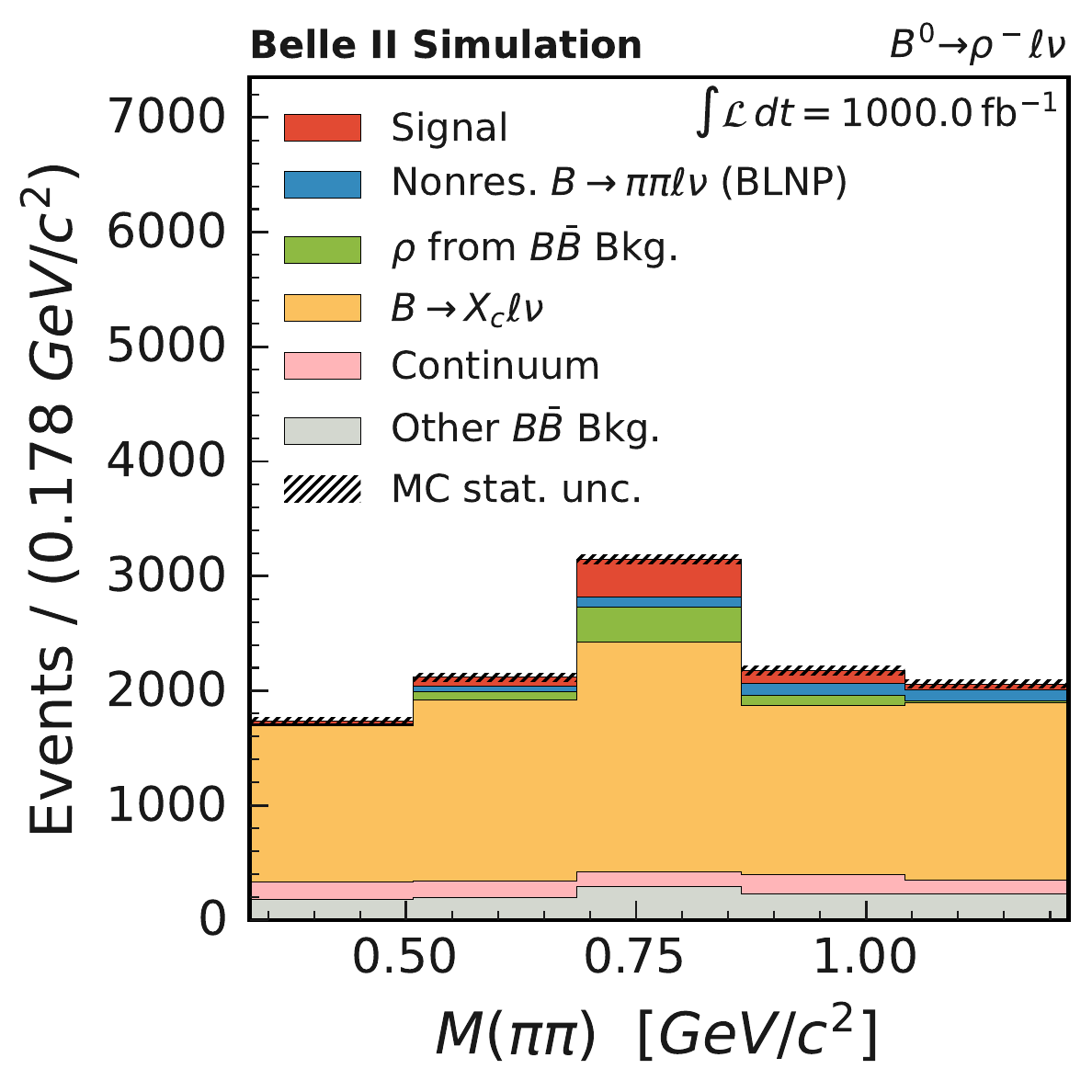}
\includegraphics[width=0.49\textwidth]{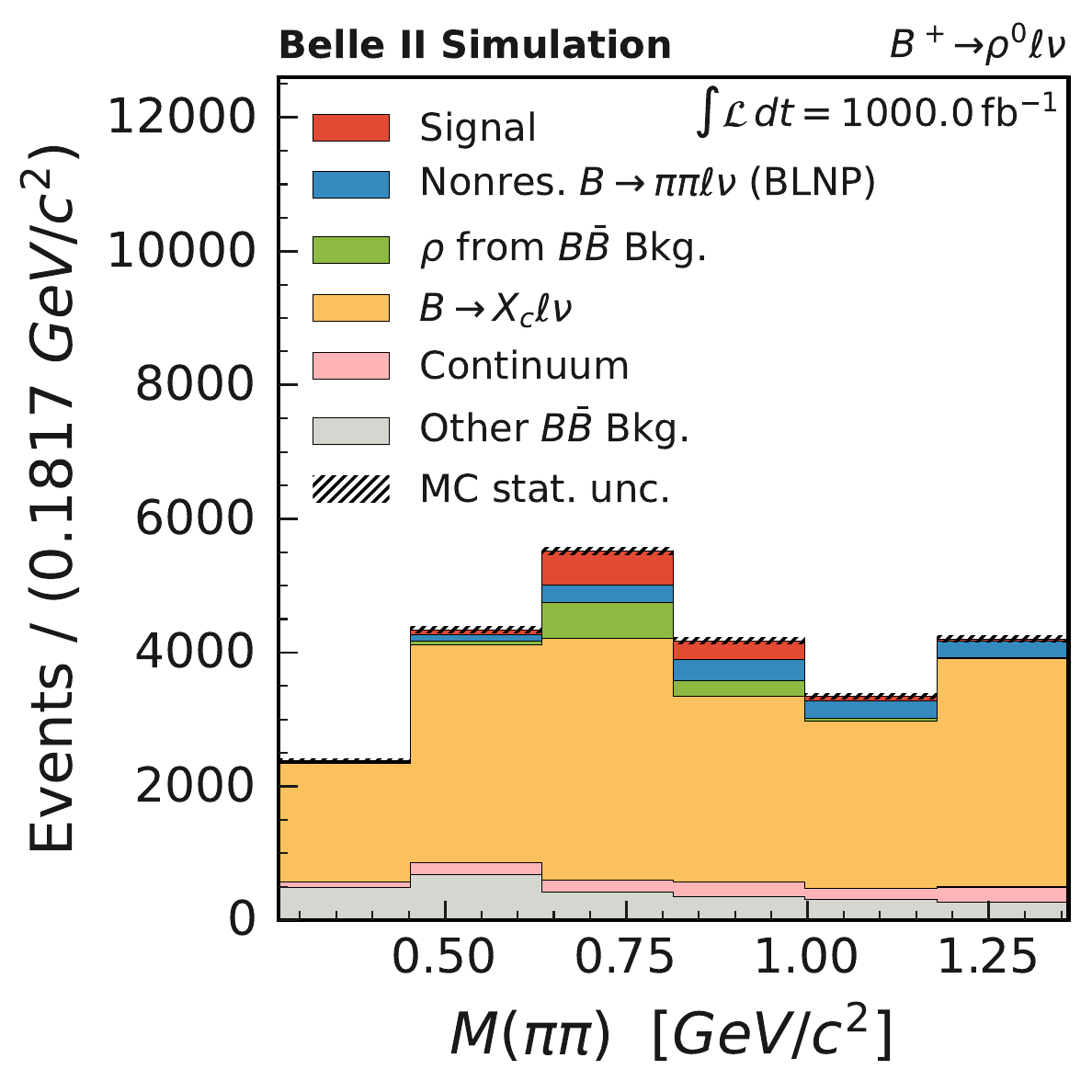}
  \caption{Projections of (top) $M_{\mathrm{miss}}^2$ and (bottom) \mpipi{} for (left) $B^0 \to \rho^-\ell^+\nu_\ell$ and (right) $B^+ \to \rho^0\ell^+\nu_\ell$ candidates reconstructed from a sample corresponding to 1\invab of simulated data. The branching fractions of the signal components are normalized to the world averages from Ref. \cite{Zyla:2020zbs}.}
  \label{fig:prefitrholnu}
\end{center}
\end{figure}

The resultant $M_{\mathrm{miss}}^2$ and \mpipi{} distributions in the MC simulation are displayed in \Cref{fig:prefitrholnu}. The MC sample is separated into distinct components to illustrate the relative contributions of various background processes. 
The background distributions shown in \Cref{fig:prefitrholnu} include the cross-feeds from nonresonant $B^0 \to \pi \pi \ell^+ \nu_\ell$ decays, $\rho$ mesons from other $B\bar{B}$ decays and a large contribution from $B \to X_c \ell \nu_\ell$ decays as well as candidates reconstructed from other generic $B\bar{B}$ and continuum events.

\section{Corrections}\label{sec:corr}
To accurately reproduce experimental data, a number of corrections are applied to the simulated data displayed in \Cref{fig:prefitrholnu}. This is necessary because both the distributions used in the fit as well as the efficiencies used in the branching-fraction calculation are derived from simulated data.

\begin{figure}[h!]
\begin{center}
	\includegraphics[width=0.49\textwidth]{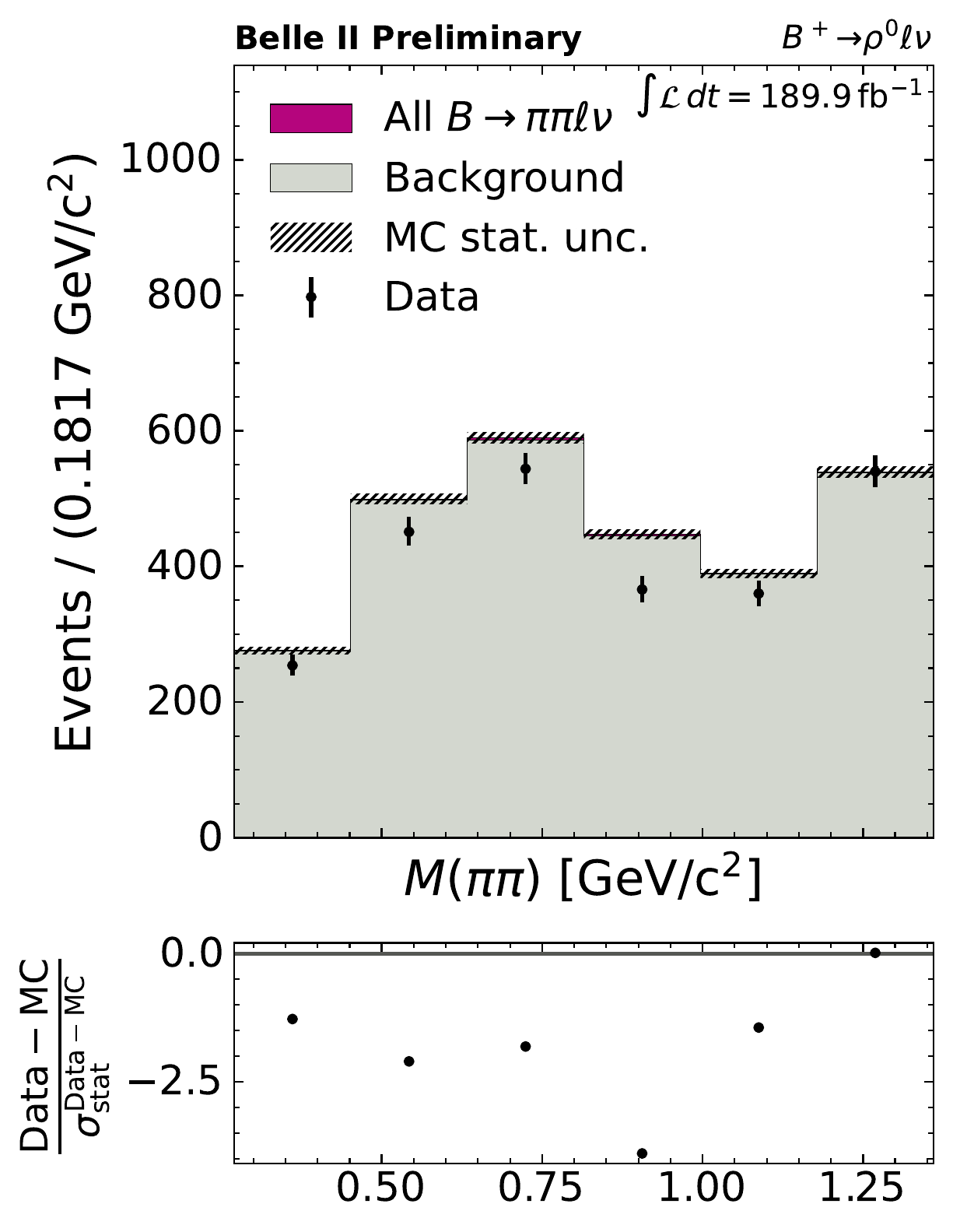}
	\includegraphics[width=0.49\textwidth]{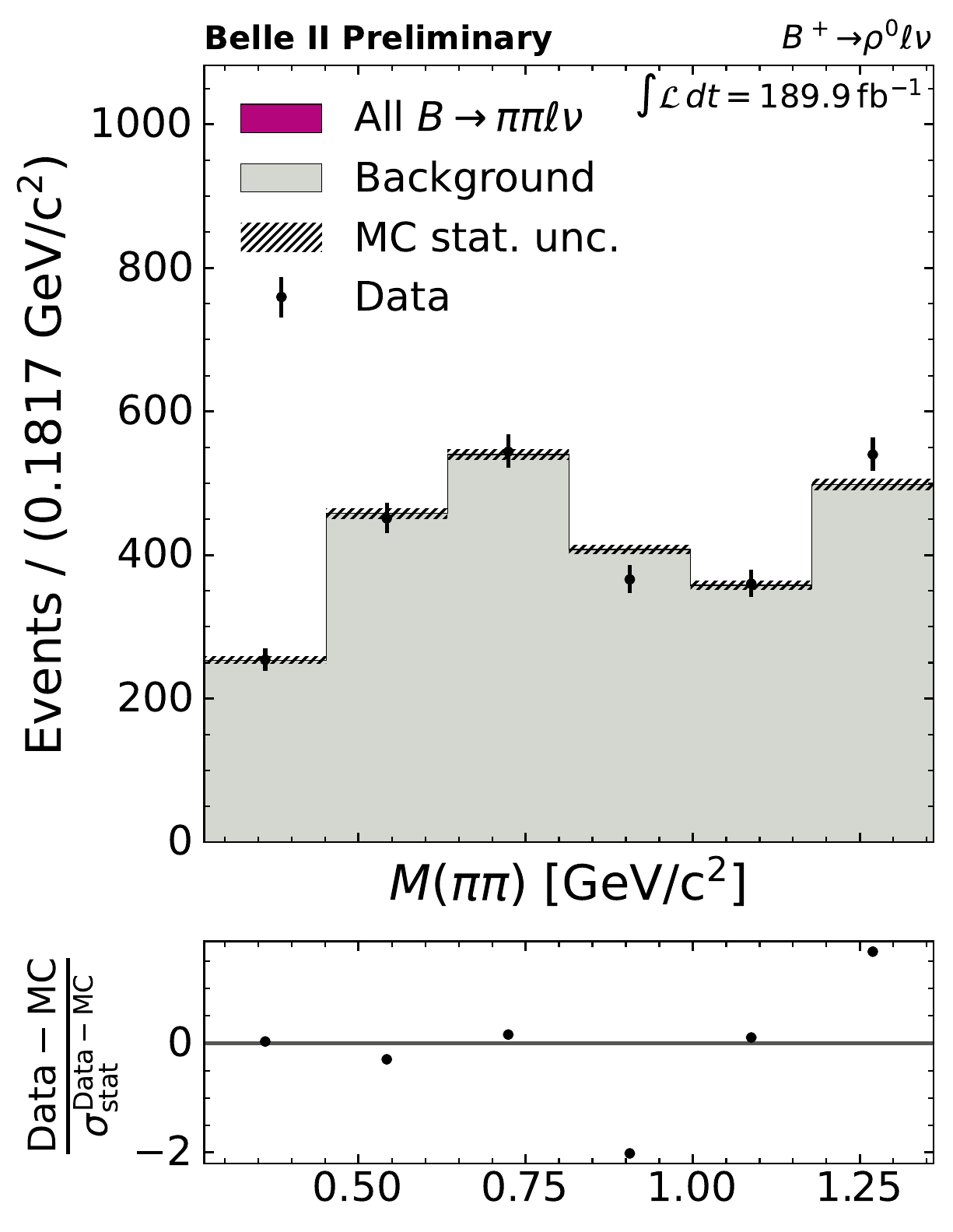}
    \caption{Distribution of \mpipi{} in the \mmisssq{} sideband ($> \SI{1.0}{\GeV\squared\per\clight\tothe{4}}$) for \Btorhozero{} candidates with projections of the simulated templates overlaid (left) before and (right) after the fit. The bottom subpanel shows the difference between the histograms derived from experimental and simulated data normalized by the statistical uncertainty of the difference between experimental and simulated data.}
  \label{fig:sidebandfit}
\end{center}
\end{figure}

The total number of MC events is scaled down by hadronic FEI calibration factors of $0.713 \pm 0.019$ for events reconstructed using neutral $B$-meson tags and $0.640 \pm 0.039$ for those reconstructed via charged $B$-meson tags, based on independent studies on control channels~\cite{Sutcliffe:2020cn}.
To test the applicability of these factors, which account for the differences between data and MC simulation in the \Btag{} reconstruction efficiency, the number of background events in data and MC simulation is compared in the sideband region $\SI{1.0}{\GeV\per\clight\squared} < \mmisssq{}$. For neutral $B$-meson tags, this ratio agrees well with the one obtained in the independent study. In charged $B$-meson tags, this ratio is not consistent with the value obtained from the lepton-momentum spectrum study. A complementary calibration factor of $0.601\pm 0.012 (\mathrm{stat})$ is obtained from the sideband region using a fit in \mpipi{} as illustrated in \Cref{fig:sidebandfit}.
In this fit, a single template is used for both resonant and nonresonant \HepProcess{\PB\to\Ppi\Ppi\Plepton\Pnulepton} processes, owing to the small fraction ($\approx 0.3\%$) of these decays in the data set. As the normalization of this template would assume negative values in the fit, it is set to zero. A second template is constructed from simulated backgrounds, originating mostly from \HepProcess{\PB\to\PXc\Plepton\Pnulepton} processes. The calibration factor obtained in this fit as well as the calibration factor from the independent inclusive study are averaged and a systematic uncertainty is assigned that covers both central values. 

For $B^0 \to \rho^-\ell^+\nu_\ell$ decays, an additional scaling factor SF$_{\pi^0} = 0.945 \pm 0.041$ is applied to the total MC yield to correct for differences in the $\pi^0$ reconstruction efficiency between MC simulation and data. This factor is determined via an independent study of \HepProcess{\Peta\to\Ppizero\Ppizero\Ppizero} decays, in which the ratio of signal efficiencies in data and MC simulation is determined through fitting the invariant mass of the \Peta{} meson.

Furthermore, each MC component is weighted by a set of corrections to account for the differences in the lepton-identification efficiencies and the pion and kaon misidentification-rates between MC and data. These corrections are obtained in an independent study \cite{LeptonID:2318} and are evaluated per track based on the magnitude of the lab-frame momentum $p$ and polar angle $\theta$ of the reconstructed lepton tracks.
A similar set of MC corrections are applied for the charged-pion identification efficiencies and the misidentification rates due to charged kaons.
To test for possible differences in the selection efficiency and distributions between simulated continuum samples and continuum background in data, the reconstruction is applied to data recorded with a beam energy \SI{60}{\MeV} below the kinematic threshold for producing $\PB{}\APB{}$ meson pairs.
All selection criteria except the requirement on the continuum-suppression classifier output are applied to this sample. A weight is then calculated from the ratio of the remaining number of events from this selection in data and MC and applied to the continuum MC sample.

\section{Signal Extraction}
\label{sec:sigextr}

The $M_{\mathrm{miss}}^2$ distribution separates efficiently between signal and all but one background component, the nonresonant $B^0 \to \pi \pi \ell^+ \nu_\ell$ background.
Both the signal and this component peak at $M_{\mathrm{miss}}^2$ = 0 because all final-state particles except the neutrino are reconstructed.
We use a second variable, \mpipi{}, which allows the separation of resonant and nonresonant $B^0 \to \pi \pi \ell^+ \nu_\ell$ decays, as the resonant signal produces a narrower structure than the nonresonant decay. 

\begin{figure}[h!]
\begin{center}
	\includegraphics[width=0.49\textwidth]{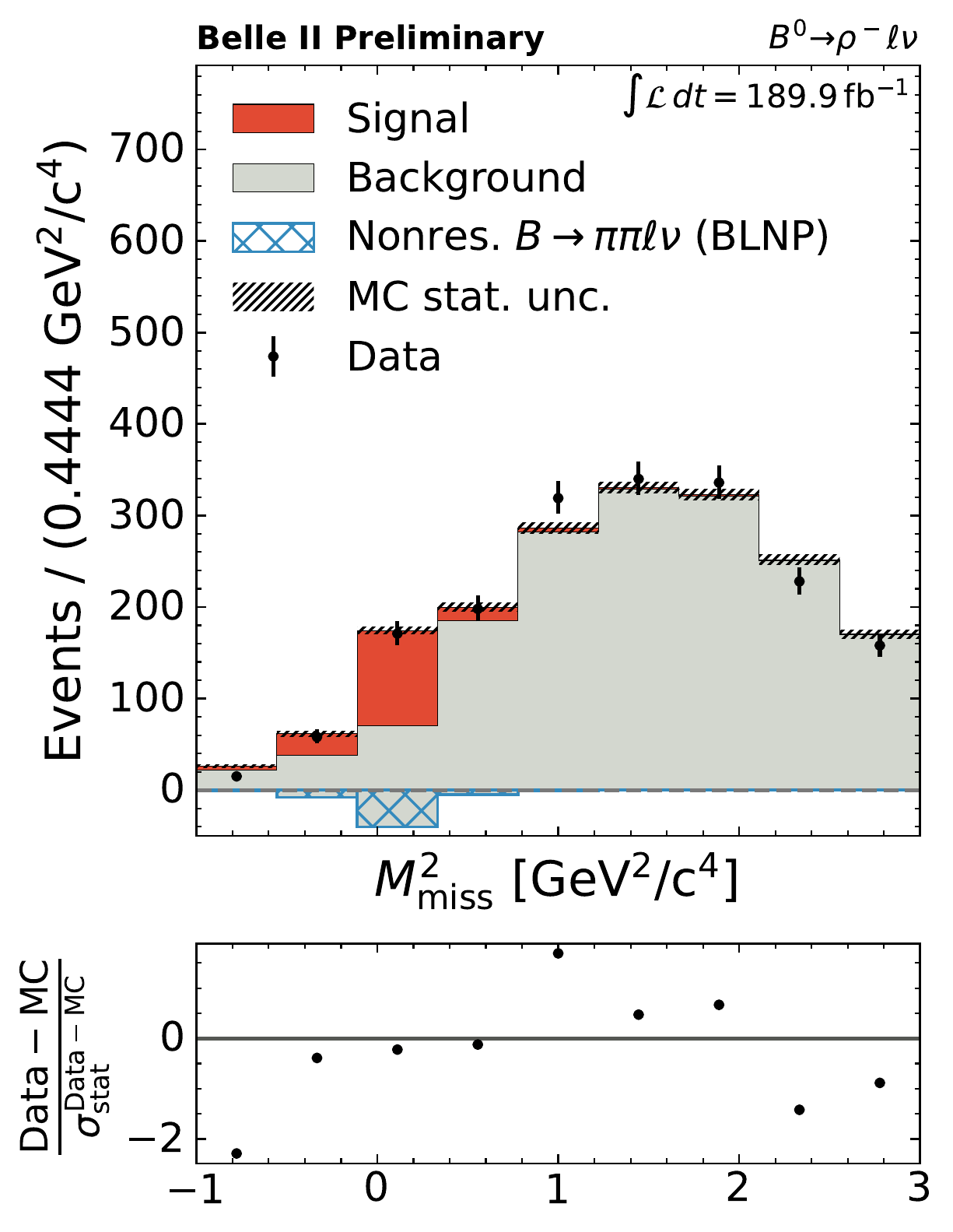}
	\includegraphics[width=0.49\textwidth]{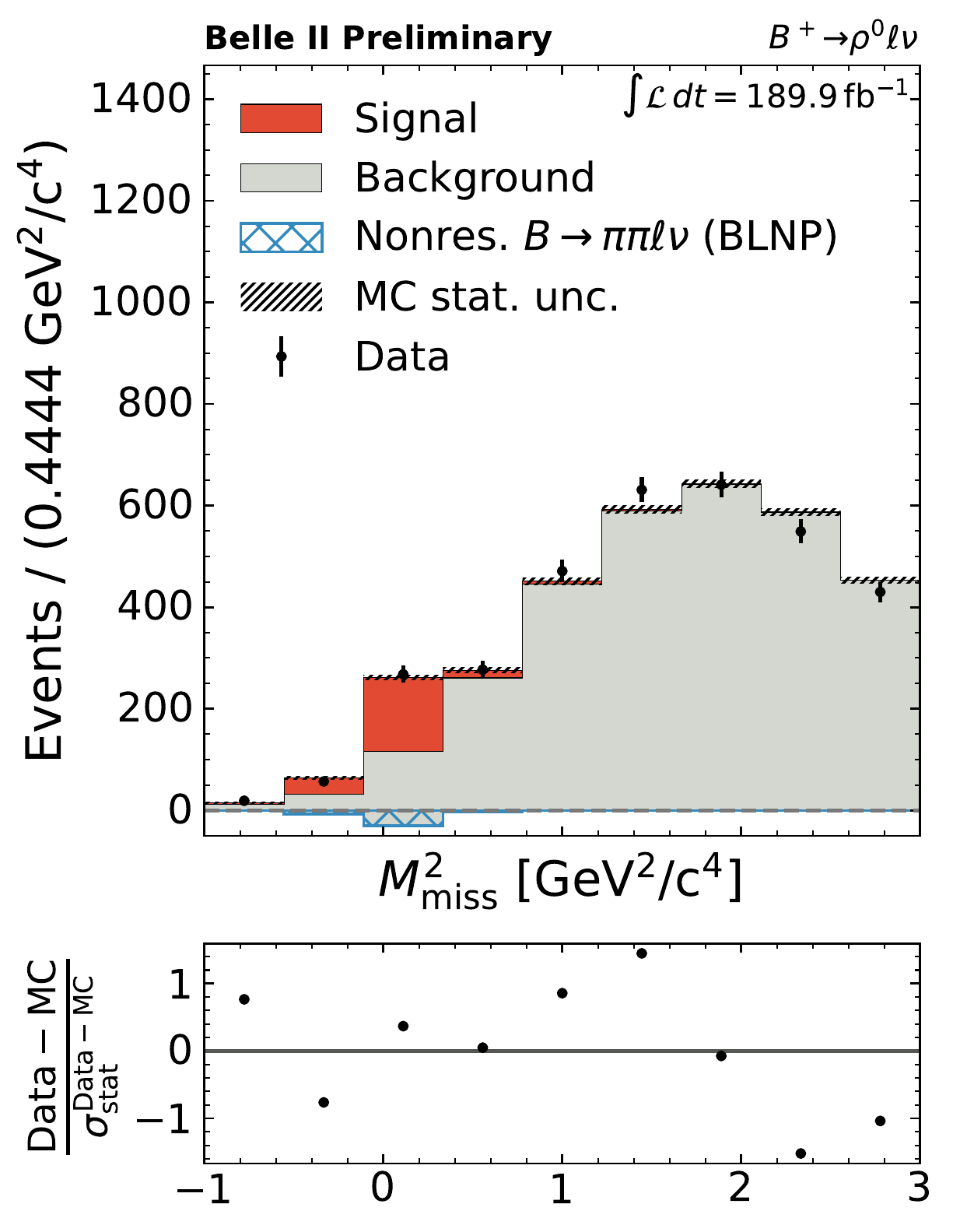}
	\includegraphics[width=0.49\textwidth]{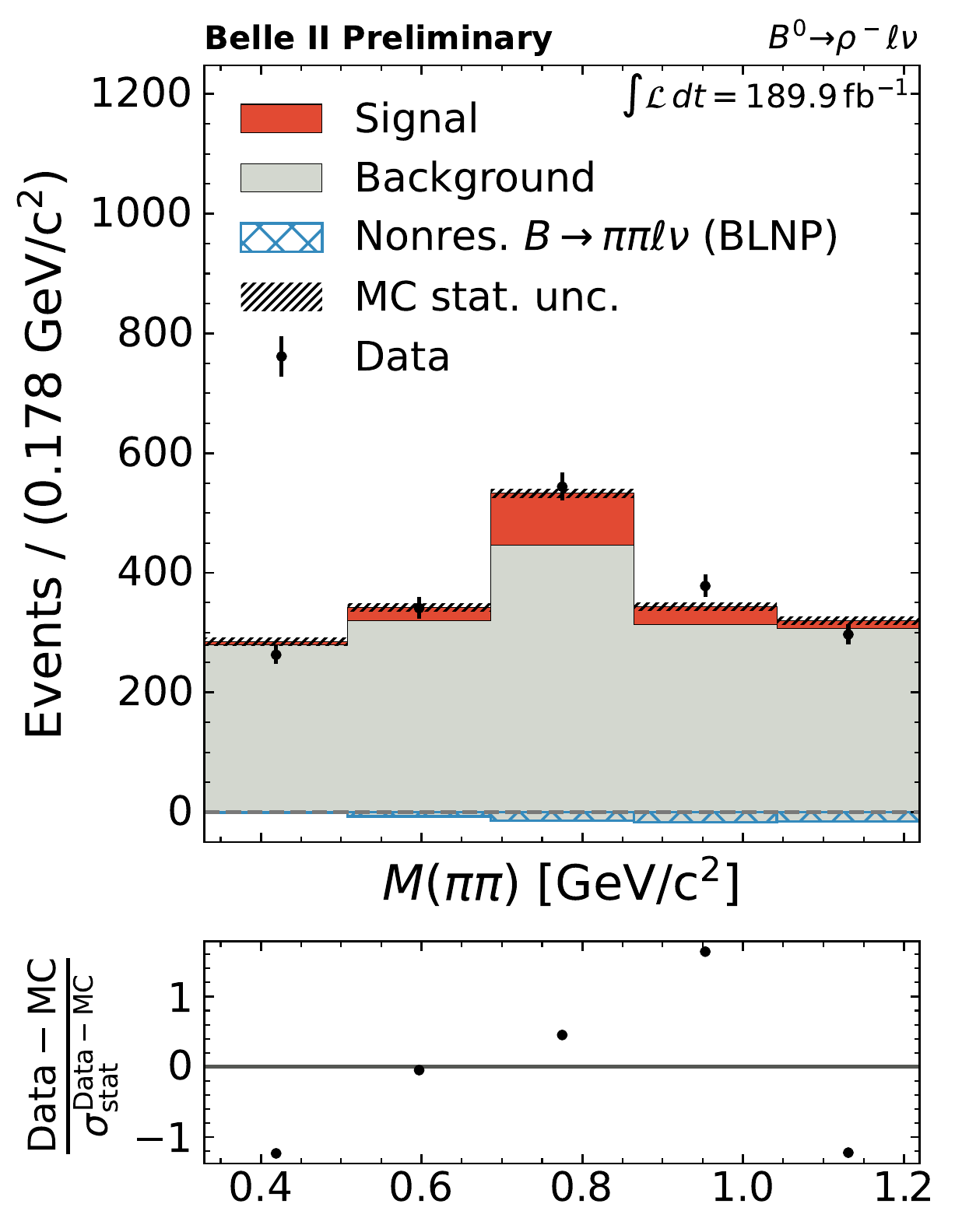}
	\includegraphics[width=0.49\textwidth]{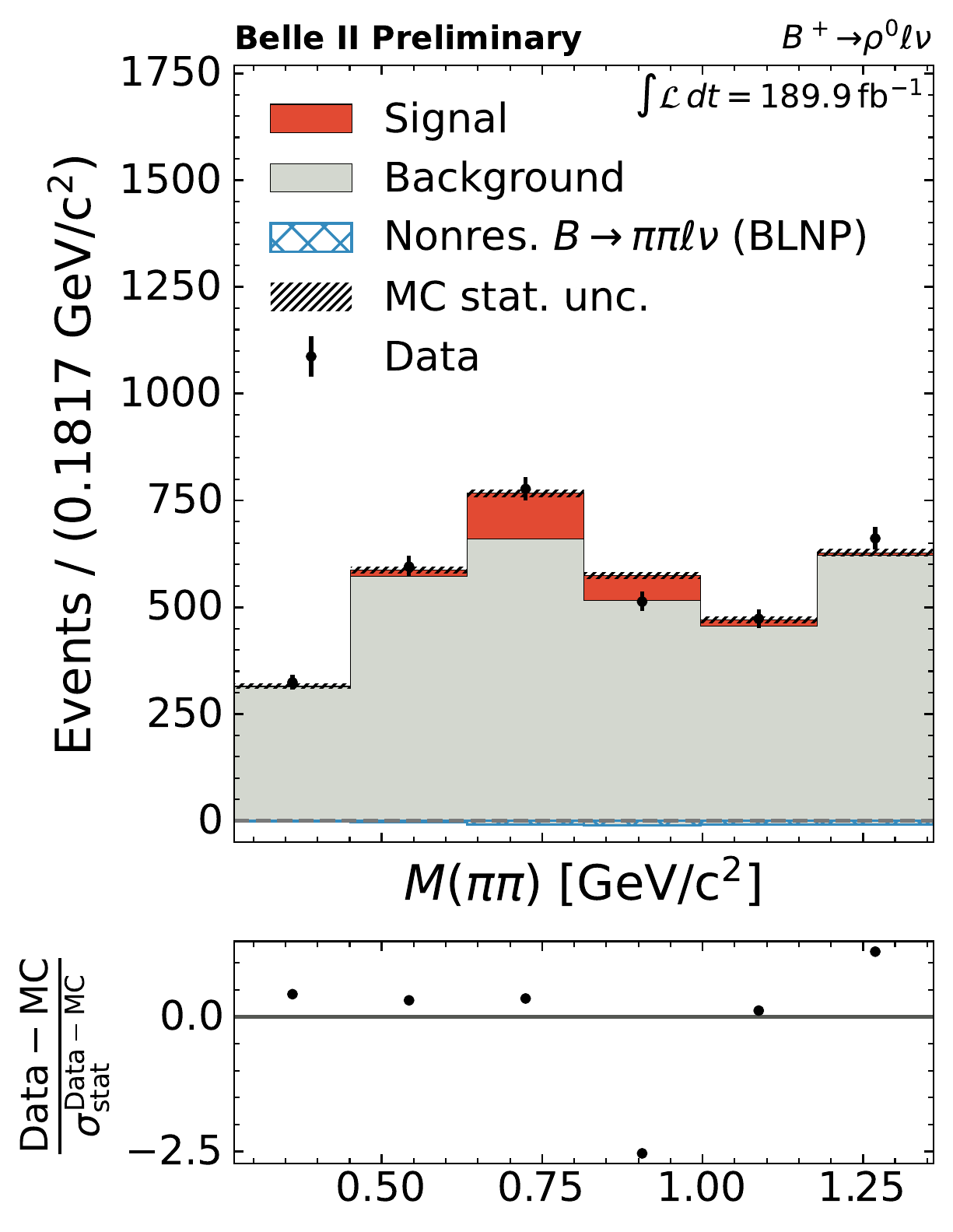}
    \caption{Distributions of (top) $M_{\mathrm{miss}}^2$ and (bottom) \mpipi{} for (left) \Btorhoplus{} and (right) \Btorhozero{} candidates reconstructed from a sample corresponding to \lint of experimental data with fit projections overlaid. The bottom subpanels show the difference between experimental and simulated data normalized to the statistical uncertainty of the difference between experimental and simulated data.}
    \label{fig:postfitrholnu}
\end{center}
\end{figure}

Using simulated samples, three template probability density functions (PDFs) for the signal component, a possible nonresonant \HepProcess{\PB\to\Ppi\Ppi\Plepton\Pnulepton} component, and all other backgrounds are constructed. 
To account for the limited sample size of the MC samples from which the templates are built, nuisance parameters corresponding to the number of simulated events in each histogram bin are introduced for each template and each bin in the fit model.
A two-dimensional, three-component, extended binned maximum likelihood fit is then performed to the \mmisssq{} and \mpipi{} distributions in data under the signal-plus-background hypothesis. The fit returns three main parameters, the signal yield and yields for both background components. All yields and nuisance parameters are allowed to float in the fit. Projections of these two-dimensional PDFs to one dimension are shown in \Cref{fig:postfitrholnu}. 

Additional extended maximum likelihood fits are then performed on the same data samples under the background-only hypothesis. The likelihood ratio $\lambda$ between fits is computed for each decay mode, $\lambda = \mathcal{L}_{S+B}/\mathcal{L}_{B} \hspace{0.5em},$ where $\mathcal{L}_{S+B}$ and $\mathcal{L}_{B}$ are the maximized extended likelihoods returned by the fits to the signal-plus-background and background-only hypotheses, respectively.
A significance estimator $\Sigma$ is defined based on the likelihood ratio, $\Sigma = \sqrt{2\ln\lambda}\hspace{0.5em}$. The fitted yields for each decay mode are listed in Table \ref{table:datafitresults}, together with the observed significances. The best fit parameters found in both fits correspond to negative yields for the nonresonant components as no bounds are placed on the normalization parameters while the bin-by-bin nuisance parameters still allow the minimization to converge. A systematic uncertainty is assigned for this effect by repeating the fit without the nonresonant component.

The branching fractions for each decay mode are extracted using the following formulae:\\

\begin{equation}
\mathcal{B}(B^0 \to \rho^- \ell^+ \nu_\ell) = \frac{N_{\mathrm{sig}}^{\mathrm{data}}(1 + f_{\mathrm{+0}})}{4\: \epsilon \:  \mathrm{CF}_{\mathrm{FEI}} ( \mathrm{SF}_{\pi^0})   \:  N_{B\bar{B}} } \hspace{0.5em},
\end{equation}
\begin{equation}
\mathcal{B}(B^+ \to \rho^0 \ell^+ \nu_\ell) = \frac{N_{\mathrm{sig}}^{\mathrm{data}}(1 + f_{\mathrm{+0}})}{4 \:  \epsilon  \: \mathrm{CF}_{\mathrm{FEI}}  \:  f_{\mathrm{+0}} \:  N_{B\bar{B}} } \hspace{0.5em},
\end{equation}
where $N_{\mathrm{sig}}^{\mathrm{data}}$ is the fitted signal yield, $f_{\mathrm{+0}}$ is the ratio between the branching fractions of the decays of the $\Upsilon$(4S) meson to pairs of charged and neutral \PB{} mesons \cite{Zyla:2020zbs}, $\mathrm{CF}_{\mathrm{FEI}}$ is the FEI calibration factor between data and MC simulation, $\mathrm{SF}_{\pi^0}$ is the scaling factor to correct the $\pi^0$ reconstruction efficiency, $N_{B\bar{B}}$ is the number of $B$-meson pairs determined in the current data set, and $\epsilon$ is the signal efficiency, which includes both electrons and muons. The factor of four in the denominator accounts for the presence of two \PB{} mesons in the $\Upsilon$(4S) decay and the reconstruction of both light lepton flavors.
The signal efficiency includes acceptance and particle identification efficiencies but does not include $\mathrm{CF}_{\mathrm{FEI}}$ and $\mathrm{SF}_{\pi^0}$. The efficiency is calculated from the fraction of generated signal events that are reconstructed after all analysis selections. 

The values of the above parameters together with the measured branching fractions are summarized in \Cref{table:BFrho}.

\begin{table}
\def\arraystretch{1.2}%
\setlength{\tabcolsep}{20pt}
  \caption{Fit results. Signal yields are listed as $N_\text{sig}$, nonresonant background yields as $N_\text{NR}$ and other background yields as $N_\text{bkg}$. Observed signal significances are also listed as $\Sigma_\text{obs}$.}\label{table:datafitresults}
\begin{tabular}{ccccc}
\hline
  Decay mode & $N_\text{sig}$ & $N_\text{NR}$ & $N_\text{bkg}$ & $\Sigma_\text{obs}$ \\\hline

  $B^0 \to \rho^- \ell \nu_\ell$ & $157 \pm 24$                                        & $-56 \pm 20$ & $1973 \pm 50$   & $7.0\sigma$\\

  $B^+ \to \rho^0 \ell \nu_\ell$ & $205 \pm 27$                                         & $-42 \pm 25$ & $3817\pm 70$ & $8.2\sigma$\\
  \hline\hline
\end{tabular}
\end{table}

\begin{table}
	\centering
\caption{Measured branching fractions of $B^0 \to \rho^- \ell^+ \nu_\ell$ and $B^+ \to \rho^0 \ell^+ \nu_\ell$ decays using \lint of data. The values of the external inputs used in the measurement are also given.}\label{table:BFrho}
\begin{tabular}{lcc}
	\hline
                                             & $B^0 \to \rho^- \ell^+ \nu_\ell$                                                            & $B^+ \to \rho^0 \ell^+ \nu_\ell$\\
																		 \hline
  $N_{\mathrm{sig}}^{\mathrm{data}}$         & $157 \pm 24$                                                                           & $205 \pm 27$ \\
 $f_{\mathrm{+0}}$                           & \multicolumn{2}{c}{1.058 $\pm$ 0.024}                                               \\
 $\mathrm{CF}_{\mathrm{FEI}}$                & $0.713 \pm 0.019$                                                                      & $0.640 \pm 0.039 $\\
  $\mathrm{SF}_{\pi^0}$                      & $0.945 \pm 0.004 \pm 0.041$                                                            & --- \\
 $N_{B\bar{B}}$                              & \multicolumn{2}{c}{(198.0 $\pm$ 3.0) $\times 10^6$}                                \\
 $\epsilon$                                  & ($0.147 \pm 0.007$)\%                                                                  & ($0.445 \pm 0.0015$)\% \\
 \hline
  $\mathcal{B}$                              & \resultrhocstat{}             & \resultrhozstat{}            \\
 \hline\hline
\end{tabular}
\end{table}

\section{Systematic Uncertainties}
\label{sec:systematics}

A number of sources of systematic uncertainty are identified for this analysis and their effects evaluated for the branching-fraction measurements. The fractional uncertainties are summarized in \Cref{table:syserrors} and include the following contributions:

\begin{itemize}
    \item \boldmath $f_{\mathrm{+0}}$: \unboldmath We combine the uncertainties on the world averages for the branching fractions $\mathcal{B}(\Upsilon$(4S)$\to B^+B^-$) and $\mathcal{B}(\Upsilon$(4S) $\to B^0\bar{B}^0$) and calculate the relative uncertainty on the fraction $f_{\mathrm{+0}}$ as a systematic uncertainty.
    \item \textbf{FEI calibration:} The uncertainty on the calibration factor for the hadronic FEI is determined taking into account multiple systematic effects in the fitting to the lepton-momentum spectrum of $B\to X \ell \nu_\ell$ decays~\cite{Sutcliffe:2020cn}. These include uncertainties on both the branching fractions and form factors of the various semileptonic components of $B\to X \ell \nu_\ell$, the lepton ID efficiency and misidentification-rate uncertainties, tracking uncertainties, and template uncertainties due to MC sample size.
      For charged \PB{} mesons, the observed normalization in the \mmisssq{} sideband is not consistent with the systematic uncertainty derived in the study of $B\to X \ell \nu_\ell$ decays. A systematic uncertainty is assigned that covers the central values of both the observed normalization difference and the factor from the independent study. 
    \item \textbf{$\pi^0$ efficiency:} The uncertainty on the scaling factor to correct the $\pi^0$ efficiency in MC is derived from an independent \HepProcess{\Peta\to\Ppizero\Ppizero\Ppizero} study comparing the number of reconstructed \Peta{} mesons in data and MC.
    \item \boldmath $N_{B\bar{B}}$: \unboldmath The uncertainty on the number of $B\bar{B}$ events includes systematic effects due to uncertainties on the luminosity, beam-energy spread and shift, tracking efficiency, and the efficiency for selecting $B\bar{B}$ events.
    \item \textbf{Reconstruction efficiency:} We represent the uncertainty on the reconstruction efficiency with a binomial uncertainty dependent on the size of the MC samples used for the analysis.  
    \item \textbf{Tracking efficiency:} We assign a systematic uncertainty of 0.69$\%$ for each charged particle. We assume complete correlation between the uncertainties for the lepton and pion tracks.
    \item \textbf{Lepton identification:} The lepton efficiencies and pion misidentification rates are evaluated in bins of the lepton momentum $p$ and polar angle $\theta$, each with statistical and systematic uncertainties. The effect of these uncertainties on the signal reconstruction efficiency is determined by generating 200 variations on the nominal correction weights via Gaussian smearing. The relative uncertainty is then taken from the spread of the values of the reconstruction efficiency over all variations. 
    \item \textbf{Pion identification:} The pion efficiencies and kaon misidentification rates as well as their associated uncertainties are evaluated the same way as the lepton identification corrections. 
    \item \textbf{$B \to X_c \ell \nu_\ell$ branching fractions:} To evaluate the effect of uncertainties in the branching fractions of several exclusive $B \to X_c \ell \nu_\ell$ decay modes, the branching fractions of the modes $B \to D^{*0} \ell \nu_\ell$, $B \to D^0 \ell \nu_\ell$, $B \to D^{*+} \ell \nu_\ell$, $B \to D^+ \ell \nu_\ell$, and $B \to D^{*} \pi^+ \pi^- \ell \nu_\ell$ are varied within their uncertainties. The branching fractions of the possible nonresonant decay modes $B \to D^{*0} \pi^+ \ell \nu_\ell$ and $B \to D^{*+} \pi^- \ell \nu_\ell$ are varied by $\pm 100\%$. For each variation, new templates are generated and the fit to data is repeated, taking the difference from the nominal result as systematic uncertainty. To combine the uncertainties from each variation, no correlation between them is assumed. This procedure is applied to the following two sources of systematic uncertainty as well. 
    \item \textbf{$B \to X_u \ell \nu_\ell$ branching fractions:} To evaluate the effect of uncertainties in the branching fractions of simulated $B \to \rho \ell \nu_\ell$, $B \to \omega \ell \nu_\ell$, and non-dipion inclusive $B \to \Xu \ell \nu_\ell$ decays in the background template, these decays are varied within their uncertainties. This corresponds to a variation of 100\% for the inclusive non-dipion decays whose rates are not known. 
    \item \textbf{$B \to f_2/f_0 \ell \nu_\ell$ branching fraction:} To evaluate the impact of these unmeasured channels on the measured $B^0 \to \rho^- \ell^+ \nu_\ell$ and $B^+ \to \rho^0 \ell^+ \nu_\ell$ branching fractions, the fit to data is repeated with the $B \to f_2/f_0 \ell \nu_\ell$ branching fraction varied by 100\% while maintaining the overall inclusive normalization via hybrid weights.
    \item \textbf{$B \to \rho \ell^+ \nu_\ell$ form factor:} To evaluate the impact of the uncertainties associated with the form factors used to describe $B \to \rho \ell^+ \nu_\ell$ decays, these factors are varied within their uncertainties and the signal templates are reweighted accordingly for each variation\cite{markus_prim_2020_3965699}.
    \item \textbf{Nonresonant $B \to \pi \pi \ell \nu_\ell$ model:} To account for the negative normalization assigned to the $B \to \pi \pi \ell \nu_\ell$ component in the fit, the normalization of this template is set to zero and the fit repeated. We take the difference between the nominal ${B \to \rho \ell^+ \nu_\ell}$ result and the result obtained without the nonresonant template as a systematic uncertainty.
    \item \textbf{Simulation sample size:} The effect of limited MC sample size used to generate the fit templates is accounted for with nuisance parameters in each bin of the template fit. Therefore no multiplicative value is assigned in~\Cref{table:syserrors}. 
\end{itemize}

The systematic uncertainties are dominated by the uncertainty assigned to the nonresonant $B \to \pi \pi \ell \nu_\ell$ model.
Previous tagged analyses of \Btorholnu{} such as the one in Ref.~\cite{Sibidanov:2013sb} do not observe negative normalizations in the fit and assign uncertainties of \SI{5}{\percent} (\PBplus{}) and \SI{1}{\percent} (\PBzero{}).
Constraints on the normalization of the nonresonant template as imposed by the inclusive measurement presented in Ref.~\cite{beleno2021} should limit the negative yield found in the fit, thereby decreasing the systematic uncertainty assigned to this simulated component directly. 
In addition, the approach used here does not consider correlations with $B \to f_2/f_0 \ell \nu_\ell$ branching fractions. Including the inclusive measurement is expected to significantly lower the overall uncertainty in the reconstruction of the \Btorhozero{} decay mode.

\begin{table}
\def\arraystretch{1.2}%
\caption{Summary of systematic uncertainties}\label{table:syserrors}

\begin{tabular}{lcc}
\hline
Source of systematic uncertainty             & $\%$ of $\mathcal{B}$($B^0\to\rho^-\ell^+\nu_\ell$) & $\%$ of $\mathcal{B}$($B^+\to\rho^0\ell^+\nu_\ell$)\\
\hline
 $f_{\mathrm{+0}}$                           & 1.2                                                 & 1.2 \\
 FEI calibration                             & 2.7                                                 & 6.1\\
 $N_{B\bar{B}}$                              & 1.5                                                 & 1.5\\
 Reconstruction efficiency $\epsilon$        & 0.5                                                 & 0.3\\
 Tracking                                    & 0.6                                                 & 0.9\\
 Lepton ID                                   & 0.7                                                 & 0.5\\
 Hadron ID                                   & 0.3                                                 & 0.6\\
  $\pi^0$ efficiency                          & 4.4                                                 & --- \\
  $B \to f_2/f_0 \ell \nu_\ell$  BF           & --                                                  & 12.1 \\
  $B \to X_u \ell \nu_\ell$  BFs             & 2.8                                                 & 4.8 \\
  $B \to X_c \ell \nu_\ell$  BFs             & 0.5                                                 & 0.5 \\
  $B \to \rho \ell^+ \nu_\ell$ form factor   & 2.7                                                 & 0.7\\
Nonres. $B \to \pi \pi \ell \nu_\ell$ model & 27.3                                                & 14.4\\
\hline
 Total                                       & 28.2                                                & 20.5 \\\hline\hline
\end{tabular}
\end{table}

\section{Results and Summary}
\label{sec:conclusions}
Including systematic uncertainties, we obtain total measured total branching fractions of ${\bfrhoc = \resultrhoc}$ and \\ ${\bfrhoz = \resultrhoz}$.
These values are compatible within one standard deviation with those determined in Ref.~\cite{Sibidanov:2013sb} but incompatible with the results reported in Ref.~\cite{BaBar:2010efp}.
Both results agree well with the current world averages of $\mathcal{B}_{\mathrm{PDG}}(B^0 \to \rho^- \ell^+ \nu_\ell) =  (2.94 \pm 0.11 \pm 0.18) \times 10^{-4}$ and $\mathcal{B}_{\mathrm{PDG}}(B^+ \to \rho^0 \ell^+ \nu_\ell) =  (1.58 \pm 0.11) \times 10^{-4}$ \cite{Zyla:2020zbs}.
 
\section*{Acknowledgments}
\label{sec:acknowledgements}
We thank the SuperKEKB group for the excellent operation of the accelerator; the KEK cryogenics group for the efficient operation of the solenoid and the KEK computer group for on-site computing support.

\clearpage
\bibliography{belle2}
\bibliographystyle{belle2-note}

\end{document}